# Relative Permittivity in the Electrical Double Layer from Nonlinear Optics

Mavis D. Boamah, Paul E. Ohno, Franz M. Geiger,* and Kenneth B. Eisenthal[2]

[1]Department of Chemistry, Northwestern University, Evanston, IL 60208, USA

[2]Department of Chemistry, Columbia University, New York, NY 10027, USA

*Corresponding author: geigerf@chem.northwestern.edu

**Abstract.** Second harmonic generation (SHG) spectroscopy has been applied to probe the fused silica/water interface at pH 7 and the uncharged $1\bar{1}02$ sapphire/water interface at pH 5.2 in contact with aqueous solutions of NaCl, NaBr, NaI, KCl, RbCl, and CsCl as low as several 10 μM. For ionic strengths up to about 0.1 mM, the SHG responses were observed to increase, reversibly for all salts surveyed, when compared to the condition of zero salt added. Further increases in the salt concentration led to monotonic decreases in the SHG response. The SHG increases followed by decreases are found to be consistent with recent reports of phase interference and phase matching in nonlinear optics. By varying the relative permittivity employed in common mean field theories used to describe electrical double layers, and by comparing our results to available literature data, we find that models recapitulating the experimental observations are ones in which 1) the relative permittivity of the diffuse layer is that of bulk water, with other possible values as low as 30, 2) the surface charge density varies with salt concentration, and 3) the charge in the Stern layer or its thickness vary with salt concentration. We also note that the experimental data exhibit sensitivity depending on whether the salt concentration is increased from low to high values or decreased from high to low values, which, however, is not borne out in the fits, at least within the current uncertainties associated with the model point estimates.

**I. Introduction.** The interaction of ions with charged surfaces can be probed in aqueous solutions using second harmonic generation (SHG) spectroscopy.[1-10] As recently shown for off-resonant conditions,[11-12] the second harmonic electric field, $E_{2\omega}$, produced at the interface may be expressed as follows:

$$E_{2\omega} \propto \chi^{(2)} E_\omega E_\omega + \chi^{(3)} E_\omega E_\omega \int_0^\infty E_{dc}(z) e^{i\Delta k_z z} dz \qquad (1)$$

Here, $\chi^{(2)}$ and $\chi^{(3)}$ are the second- and third-order susceptibility of the interface, respectively, $\Delta k_z$ is the inverse of the coherence length of the SHG process, and $E_{dc}$ is the z-(depth) dependent electric field produced by the interfacial charges, which is given by $E_{dc} = -d\Phi(z)/dz$, where $\Phi(z)$ is the electrostatic potential.[11-14] The dependence of the SHG response on the electrostatic potential has led to its use in the form of an "optical voltmeter" for probing the electrical double layer at insulator/aqueous interfaces.[1-9,15-17] Ongoing efforts focus on the development of methods for estimating interfacial electrostatic parameters from the experimental data.

For the case $\Phi(z) = \Phi(0) e^{-\kappa z}$, i.e. a potential decaying exponentially from a charged interface located at $z=0$ to zero into the bulk aqueous phase with a Debye screening length of $1/\kappa$, the total second-order response is given by[12,18]

$$\chi^{(2)}_{total} \propto \chi^{(2)} + (\chi^{(3)}_1 + i\chi^{(3)}_2)\Phi(0) \qquad (2a).$$

With $\chi^{(3)}_1$ given by $\kappa^2/(\kappa^2+(\Delta k_z)^2)$ and $\chi^{(3)}_2$ given $\kappa\Delta k_z/(\kappa^2+(\Delta k_z)^2)$, one arrives at

$$E_{2\omega} = A + B\Phi(0)\frac{\kappa}{\kappa - i\Delta k_z} \qquad (2b),$$

where $A$ is given by $\chi^{(2)}E_\omega E_\omega$ and $B$ is given by $\chi^{(3)}E_\omega E_\omega$.[19] The last factor accounts for phase matching, which was first considered in 1992.[2] More recent work has used expressions like the ones shown in eqns. 1 and 2 to account for phase matching in nonlinear optical signal generation from interfaces that are subject to interfacial potentials.[11-14]

According to eqn. 2, SHG intensity maxima or minima may be observed as $\Phi(0)$ is varied,[11,20] whereas strict increases (*resp*. decreases) in the SHG intensity with increasingly negative (*resp*. positive) values of $\Phi(0)$ are expected in the absence of phase matching. Indeed, we reported already in 2013 charge screening experiments carried out at pH 7 on fused silica using a near total internal reflection geometry that showed an SHG intensity maximum at salt concentrations slightly below 1 mM, as opposed to strictly increasing SHG signal intensities with decreasing ionic strength.[20]

Our present work shows how to account for the optical interference between the second- and third-order terms shown in eqn. 1 and 2 when analyzing experimentally obtained non-resonant SHG data for electrostatic parameters such as interfacial charge density. Our approach is similar in spirit to what we recently reported for the case of absorptive-dispersive mixing that arises for conditions of resonantly enhanced SHG or SFG data,[14] but considers only non-resonant scenarios. We also ask whether the effect depends on the choice of the optical geometry used in the experiments and polarization of the probe and signal light fields, whether it is influenced by the presence of dissolved $CO_2$ in the aqueous solutions, and how the effect might manifest itself for an oxide/water interface held at the point of zero charge ($1\bar{1}02$ sapphire at pH 5.2). Upon consideration of the answers to these questions, we find a means for estimating the solvent relative permittivity in the electrical double layer, at least as this region is probed by nonlinear optics under off-resonant conditions.

**II. Experimental.** Our experimental approaches for performing SHG studies in external and near internal reflection have been described previously.[12] Briefly, we use an 82 MHz Ti:Sapphire oscillator at 800 nm to probe the water/fused silica interface using 120 femtosecond pulses at an incident angle of 60° from the surface normal. For experiments carried out using near internal reflection on fused silica, a fused silica hemisphere (ISP Optics) is clamped leak tight onto a Teflon flow cell having an internal sample volume of 2 mL using a Viton O-ring described in detail elsewhere.[21] For the sapphire PZC experiments, a sapphire window (ISP Optics) is clamped between the O-ring and the hemisphere. For experiments carried out using the external reflection geometry, the fused silica sample (Meller Optics) is housed inside a custom-built hollow fused quartz dome clamped leak tight onto a Teflon flow cell sealed with a Viton O-ring. The incident angle is 60° for both geometries. We work under creeping flow conditions and at low shear rates that we obtain by employing peristaltic pumps set to maintain conditions of creeping flow at constant flow rates of ~1 mL/sec across the several mL internal volume of our sample flow cell.

We used NaOH and HCl for pH balancing of all the alkali halide solutions except for CsCl, where in some cases we used CsOH. NaCl was obtained from Sigma-Aldrich (Part # 746398-2.5KG, Lot # SLBK2618V and ≥99% pure) and Alfa Aesar (Lot # M08A016 and ≥99% pure), NaBr was obtained from Sigma-Aldrich (Part # 310506-100G, Lot # MKBQ8200V and ≥99% pure), NaI was obtained from Santa Cruz Biotechnology (Lot # J1515, Catalog # sc-203388A and ≥99% pure), KCl was obtained from Sigma-Aldrich (Part # 746435-500G Lot # SLBP3785V and ≥99% pure), RbCl was obtained from Aldrich (Part # R2252-50G, Lot # WXBC0662V and ≥99% pure), CsCl was obtained from Sigma-Aldrich (Part #'s C3011-25G, C3011-100GLot # SLBP4992V and ≥99% pure) and Aldrich (Part # 203025-10G and ≥99.999% pure), CsI was obtained from Aldrich (Part # 202134-25G, Lot # MKBX2413V and ≥99.9% trace metal basis), CsOH was obtained from Aldrich (Part # 516988-25G,Lot # MKBX2444V and 99.95% trace metal basis), NaOH was obtained from EMD chemicals (Lot # B0312669 941) and HCl was obtained from Fisher Scientific (Lot # 155599). The solution pH was measured for each salt concentration and shown to remain constant over the range of ionic strengths investigated here.

We maintained the input laser power measured before the sample stage at 0.46 ± 0.05 W. The polarization combination used was set to p-in/all-out, unless otherwise noted. The SHG signal was detected using a single photon counter (SR400, Stanford Research Systems) and averaged using a boxcar procedure written in IgorPro (Wavemetrics). Multiple individual adsorption isotherms were collected on multiple days and using multiple different fused silica substrates at laboratory temperature (21°C - 22°C). Laser power fluctuations were accounted for by recording the measured power reflected from an optical element within the laser line simultaneously with the SHG collection and normalizing the detected signal intensity to the square of the measured power, P. Power stability studies show that over two hours, the average power, $<P>$, is 36.69 ± 0.03 mW with drift of 0.3 ± 6 μW/s.[22]

### III. Results and Discussion.

**III.A. SHG vs [salt] Responses Show Maxima Near 1 mM at pH 7.** Figure 1A shows the SHG signal intensity obtained from the fused silica/water interface in internal reflection geometry for pH 7 and for NaCl concentrations between 10 μM and 100 mM and p-in and s-in polarized light. Maxima at $2 \times 10^{-4}$ M salt concentration are clearly observed. The SHG intensities were normalized to the average value of that maximum. Figure 1A also shows that using an external reflection geometry produced the same results, given the incident angles (60°) stated above. Figure 1B shows the response using nitrogen-purged ($CO_2$-free) water, along with water that had been equilibrated with ambient ($CO_2$-containing) air overnight. Figure 1C shows the changes in the SHG intensities observed in this low ionic strength regime are fully reversible for the variety of alkali halide salts surveyed. Taken together, Figures 1A-C indicate that the effect of SHG response maxima in the sub-mM concentration regime reported here does not depend on the particularities of the reflection geometry used in the experimental setup or the presence of dissolved $CO_2$ or carbonate, and that it is not associated with any noticeable time delay or considerable degree of irreversibility.

**III.B. SHG Maxima not Observed at Point of Zero Charge.** To investigate whether maxima are also observable at the point of zero charge (PZC), we worked with α-alumina (PZC of 5.2 for the $1\bar{1}02$ surface),[23] given that synthetic silica's low PZC (pH 2.3)[24] prevents us to access the sub-mM salt concentrations needed to probe for the SHG maxima. Fig. 2 shows that increasing the NaCl concentration in a pH 5.2 solution over the $1\bar{1}02$ α-alumina surface from $10^{-5}$ to $10^{-4}$ M coincides with ca. 5% to 8% increases in the SHG E-field, beyond which it remains invariant with further increases in NaCl concentration. This latter result is expected for a surface that is, on average, uncharged and supports the notion that the SHG maxima discussed here are produced by optical-electrostatic interactions occurring over surfaces carrying a net charge.

**III.C. Accounting for Phase Matching.** To estimate the interfacial charge density, $\sigma$, from SHG data, a model for the interfacial potential, $\Phi(0)$ is needed. In the commonly used Gouy-Chapman (GC) model, the charge in the diffuse layer is equal to the opposite of the surface charge, provided that specific ion adsorption is negligible.[25] The interfacial potential is given as follows:

$$\Phi(0) = \frac{2k_BT}{ze}\sinh^{-1}\left[\frac{\sigma}{\sqrt{8k_BT\varepsilon_0\varepsilon_r n_i}}\right] \quad (3a)$$

Here, $k_B$ is the Boltzmann constant, $T$ is the temperature, $z$ is the valence of the electrolyte, $e$ is the elementary charge, $\varepsilon_o$ is the vacuum permittivity, $\varepsilon_r$ is the relative permittivity of bulk water, and $n_i$ is the concentration of ions. Converting to molarity [mol/L] and assuming univalent electrolytes at 25°C gives:[11]

$$\Phi(0) = 0.05139 \,[V]\sinh^{-1}\left[\frac{\sigma}{0.1174\,[C\,m^{-2}\,M^{-1/2}]\sqrt{C}}\right] \quad (3b)$$

where units on the constants are given explicitly, $\Phi_0$ is in [V], $\sigma$ is in [C/m$^2$], and C is in [M]. Eqn. 2b then becomes

$$E_{2\omega} = A + B\,0.05139\,[V]\sinh^{-1}\left[\frac{\sigma}{0.1174\,[C\,m^{-2}\,M^{-1/2}]\sqrt{C}}\right]\frac{\kappa}{\kappa - i\Delta k_z} \quad (3c)$$

The magnitude of the phase matching factor becomes important for ionic strengths $<\sim 1$ mM, as the expansion of the coherent sum in the square modulus of eqn. 2b indicates:

$$I_{SHG} \propto \left|A + B\Phi(0)\frac{\kappa}{\kappa - i\Delta k_z}\right|^2 = A^2 + \left(B^2\Phi(0)^2 + 2AB\Phi(0)\right)\frac{\kappa^2}{\kappa^2 + \Delta k_z^2} \quad (4a)$$

$$\lim_{\frac{\kappa}{\kappa - i\Delta k_z} \to 1}\left(\left|A + B\Phi(0)\frac{\kappa}{\kappa - i\Delta k_z}\right|^2\right) = A^2 + \left(B^2\Phi(0)^2 + 2AB\Phi(0)\right) \quad (4b)$$

When the Debye length is long compared with the wavevector mismatch (I ≤ 1 mM), the electrostatic field penetrates far enough into the aqueous medium for dispersion between the fundamental and second harmonic beams to become significant, and the phase correction term approaches and then exceeds 10% (Table I). When the Debye length is short compared with the wavevector mismatch (I ≥ 1 mM), the generated signal intensity is produced close to the interface, minimizing the effect of dispersion, and the phase correction term nears unity. Indeed, earlier work has shown that for electrolyte concentrations >1 mM, the Gouy-Chapman, the constant capacitance, and even model-independent treatments give excellent agreement with experiment without any need for ionic strength-dependent phase corrections.[1-9,15-17]

For simplicity, one could fit the SHG dataset in the high-salt concentration regime only, say, between 1 mM and higher, assuming $\frac{\kappa^2}{\kappa^2+\Delta k_z^2} = 1$. However, given the additional information provided by the data points collected in the salt concentration regime, we offer the following data analysis: inserting the GC model into eqn. 2b and rearranging, we obtain

$$(E_{SHG} - A)/\frac{\kappa}{\kappa - i\Delta k_z} = B\left\{0.05139 \text{ arcsinh}\left(\frac{\sigma}{0.1174\sqrt{C}}\right)\right\} \quad (5)$$

Taking the A parameter in eqn. 5 from the square root of the average of the measured highest-concentration SHG intensity data points in Fig. 1B, (A=0.54 ± 0.07), we use the Mathematica notebook provided in the Supporting Information to replot the SHG intensities shown in Figure 1A in the form of Figure 3A. The SHG responses, once accounting for the phase matching term, first decrease monotonically with increasing salt concentration and then approach zero for salt concentrations exceeding *ca.* 10 mM under the conditions of our experiment. Fitting eqn. 5 to the SHG dataset results in a charge density of -0.013 ± 0.001 C/m$^2$ and a B value of 4.2 ± 0.1 when A=0.54. For A=0.47 and 0.61, estimates of the lower (*resp*. upper) bounds of the charge density and B are -0.011 ± 0.001 C/m$^2$ (*resp*. -0.015 ± 0.002 C/m$^2$) and -5.1 ± 0.1 (*resp*. -3.2 ± 0.1).

**III.D. Model Sensitivities.** The charge density point estimates obtained from the SHG charge screening analysis presented here are in good agreement with those compiled by Hore and co-workers[26] from electrokinetic measurements carried out in the 1 to 10 mM salt concentration regime. Yet, even when we restrict our analysis to fitting eqn. 5 to ionic strengths ranging from the lowest salt concentration to increasingly larger concentrations, charge density point estimates are three to four times smaller than those reported from electrokinetic measurements carried out above 10 mM salt concentration (Fig. 3B, similar fit results are obtained when using the x-axis range limits in reverse). We therefore proceeded to examine some of the model sensitivities, as outlined next. Specifically, given that the other parameters in the model are fundamental physical constants ($k_B$, $T$, $e$, $N_A$, and $\varepsilon_o$), or held constant during the experiment ($z$, $E\omega$), we examined the sensitivity of eqn. 5) to possible variations in 1) charge density, $\sigma$, with ionic strength, and 2) departures of the relative permittivity, $\varepsilon_r$, from its value of 80 in the bulk at room temperature. The first consideration is motivated by reports of salt-dependent surface charged densities in the 1 mM to 1 M salt concentration regime at pH 7,[27-29] which generally show somewhat elevated charged densities with elevated ionic strength. The second consideration seems to be reasonable to entertain given that relative permittivities having values significantly below 80 have been reported from theoretical and computational analyses,[30-36] at least for distances close (<10 nm) to the interface. We note that variations in $\chi^{(2)}$ and/or $\chi^{(3)}$ with salt have been reported to be minor.[4,13]

To examine the possibility of a salt concentration-dependent surface charge density, we replaced $\sigma$ in eqn. 3c with an empirical equation of the form $\sigma = -0.041\pm0.006 + 0.032\pm0.006 \cdot e^{(-3\pm1 I)}$, which we obtained from fitting experimental interfacial charge densities reported by Ahmed,[27] Abendroth,[28] and Kitamura et al.,[29] again, as compiled by Hore and co-workers.[26] Using this expression, Figure 3B shows good agreement between two models (the B parameter for this fit is $-4.65 \pm 0.03$).

To test whether the presence of higher salt concentrations can lead to more negative charge densities in SHG charge screening experiments, we exposed a fresh fused silica surface to a pH 7 solution maintained at 300 mM salt concentration, followed by charge screening beginning at 1 mM and going back to 300 mM salt (Figure 4). The SHG signal intensities are up to 10-20% above those obtained when first increasing the salt concentration from a few 10s of μM salt to several 100 mM salt over a fused silica surface that had never been exposed to high salt, and then reducing the salt concentration back to a few 10s of μM. Indeed, at 0.1 mM salt concentration, the SHG signal intensity differences are on average, from two independent replicates, 380±18 vs 440±32 counts per second when increasing vs decreasing the salt concentration in the charge screening experiments. Unfortunately, GC model fits to either branch (low-to-high salt or reverse) of this dataset reveal no statistically significant differences in the charge density point estimates ($-0.017\pm2$ C/m$^2$ vs $-0.014\pm1$ C/m$^2$).

We then proceeded to examine $\varepsilon_r$, which appears both in the GC model as well as the phase matching term in eqn. 5, using relative permittivities that varied between 1 and 80. For this analysis, we transformed the SHG intensity data according to eqn. 4a, using various values for the relative permittivity in the correction term. We then fit the GC model using those same relative permittivities to obtain B and σ, which are summarized in Figure 5.

The charge density point estimates obtained from the GC fits are within the range typically reported for fused silica at circumneutral pH and comparable ionic strengths[26,37] for values of $\varepsilon_r > 30$. While relative permittivities << 80 have been reported for the first few nm or so of the interfacial region,[30-36] it is unclear why the relative permittivity should also be low in deeper probe regions of the SHG process, which in our case can be many tens and even hundreds of nanometers, depending on the ionic strength.[11] Therefore, our analysis appears to support the notion that the relative permittivity relevant for the SHG process described here is well represented by a value corresponding to that of bulk water. Note that the frequently used Gouy-Chapman-Stern (GCS) model, in which another parameter, the $C_{Stern}$ capacitance, is introduced according to

$$(E_{SHG} - A) / \frac{\kappa}{\kappa - i\Delta k_z} = B \left\{ \frac{-\sigma}{C_2} + \frac{2kBT}{ze} \text{arcsinh} \left( \frac{\sigma}{\sqrt{8000 k_B T N_A \varepsilon_0 \varepsilon_r C}} \right) \right\} \quad (6)$$

produces good agreement when, instead of applying the commonly used[38] $C_2$ value of 0.2 F/m$^2$, one uses a $C_2$ value of 0.8 (s=-0.025 ± 7 C/m$^2$ and B=17 ± 1), with the fit parameters quickly converging to -0.014 ± 1 C/m$^2$ and B=17 ± 1 for larger values of $C_2$. and 2.0 along with a relative permittivity of 80 for $\varepsilon_r$ (data not shown). This seems to indicate a Stern layer as thin as 5 Å, in reasonable agreement with recently reported work by Brown *et al.*[39] We note that the expression $C_2 = \varepsilon_o \varepsilon_{Stern}/\delta$, which is used to estimate the Stern layer thickness in that particular work, assumes a value of 43 for the relative permittivity in the Stern layer, $\varepsilon_{Stern}$.[40]

**V. Conclusions.** In conclusion, we have probed the charged fused silica/water interface at pH 7 and the uncharged $1\bar{1}02$ sapphire/water interface at pH 5.2 using SHG spectroscopy and using aqueous solutions of NaCl, NaBr, NaI, KCl, RbCl, and CsCl as low as several 10 µM. Two response regimes were found in our experiments. For ionic strengths up to about 0.1 mM, the SHG responses were observed to increase, reversibly for all salts surveyed, when compared to the condition of zero salt added. Further increases in the salt concentration led to monotonic decreases in the SHG response. The SHG increases followed by decreases were found to be consistent with recent reports of phase interference and phase matching in nonlinear optics. By varying the relative permittivity employed in common mean field theories used to describe electrical double layers, and by comparing our results to available literature data, we find that the model that recapitulates the experimental observations is one in which 1) the relative permittivity of the diffuse layer is that of bulk water, with other values as low as 30 possible, 2) the surface charge density varies with salt concentration, and 3) the charge in the Stern layer may also depend on salt concentration, if using the Gouy-Chapman-Stern model. We also note that the experimental data exhibit sensitivity depending on whether the salt concentration is increased from low to high values or decreased from high to low values, which, however, is not borne out in the fits, at least within the current uncertainties associated with the model point estimates.

**Supplementary Material.** See the supplementary material for the Mathematica notebook used for the analysis presented in this work.

**Acknowledgments.** This work was supported by the US National Science Foundation (NSF) under its graduate fellowship research program (GRFP) award to PEO. KBE and FMG gratefully acknowledge NSF award numbers CHE-1057483 and CHE-1464916, respectively.

__

_

**Author Information.** The authors declare no competing financial interests.

Correspondence should be addressed to FMG (geigerf@chem.northwestern.edu).

**Tables and Table Captions.**

**Table 1.** Correction factor magnitude at select ionic strengths and incident angles for 800 nm fundamental beam and the fused silica/water interface.

| Ionic Strength [mM] | $\dfrac{\kappa^2}{\kappa^2 + \Delta k_z^{\,2}}$ | | |
|---|---|---|---|
| | 45° | 60° | 70° |
| 0.05 | 0.30 | 0.35 | 0.38 |
| 0.1 | 0.46 | 0.51 | 0.55 |
| 1 | 0.90 | 0.91 | 0.92 |
| 10 | 0.99 | 0.99 | 0.99 |
| 100 | 1.0 | 1.0 | 1.0 |

Boamah et al. Page 17

**Figures and Captions**

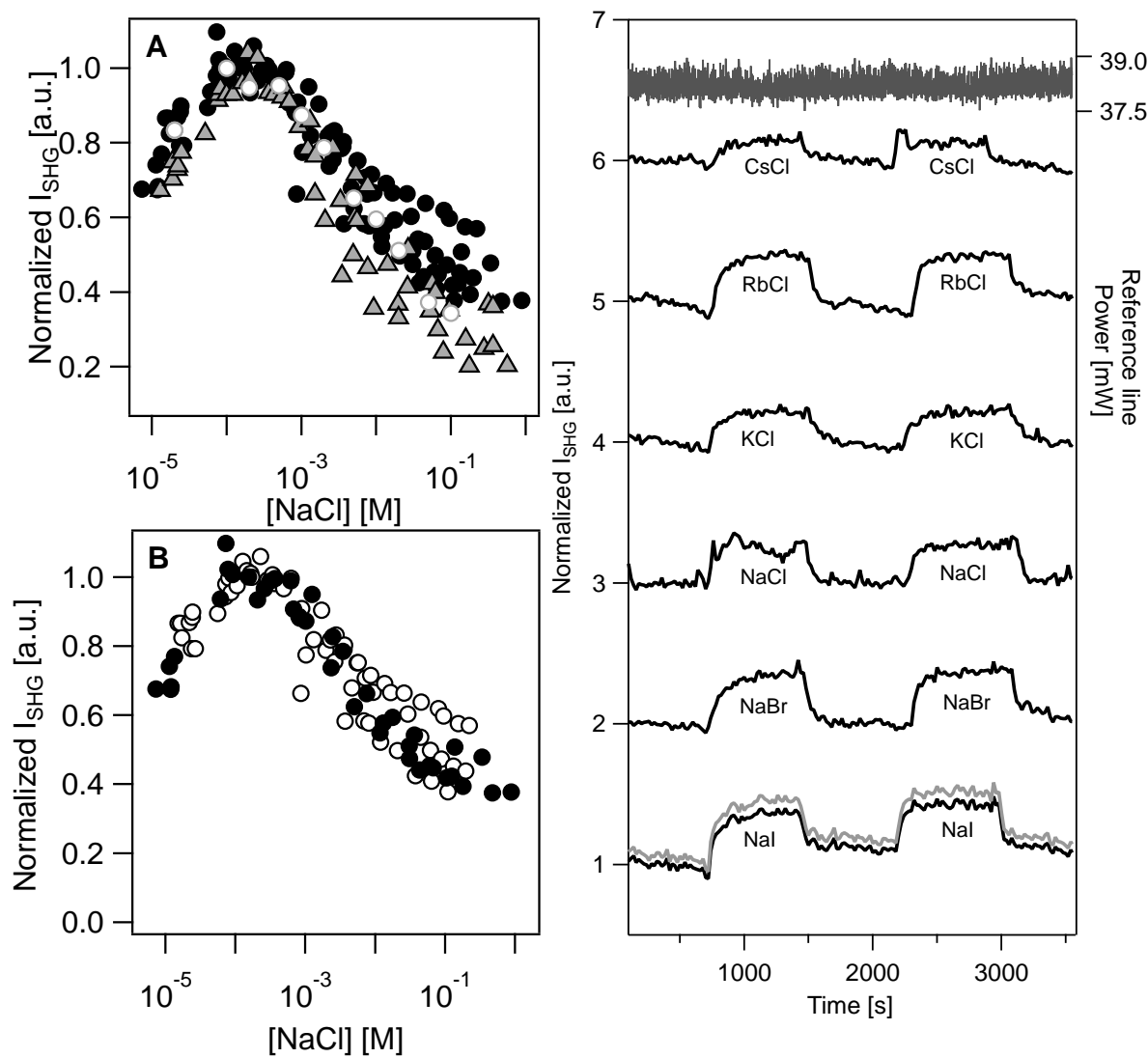

**Figure 1.** (**A**) SHG responses observed for the fused silica/water interface held at pH 7 and subjected to various salt concentrations under conditions of external (empty circles, p-in polarization) and internal (filled circles for s-in polarization and filled triangles for p-in polarization) reflection geometry. (**B**) Same as (**A**) but for nitrogen-purged (filled circles) and unpurged (empty circles) water. (**C**) SHG intensity vs. time traces for various salts studied in the sub-mM concentration regime surveyed here along with the readout of the laser power output (right y-axis).

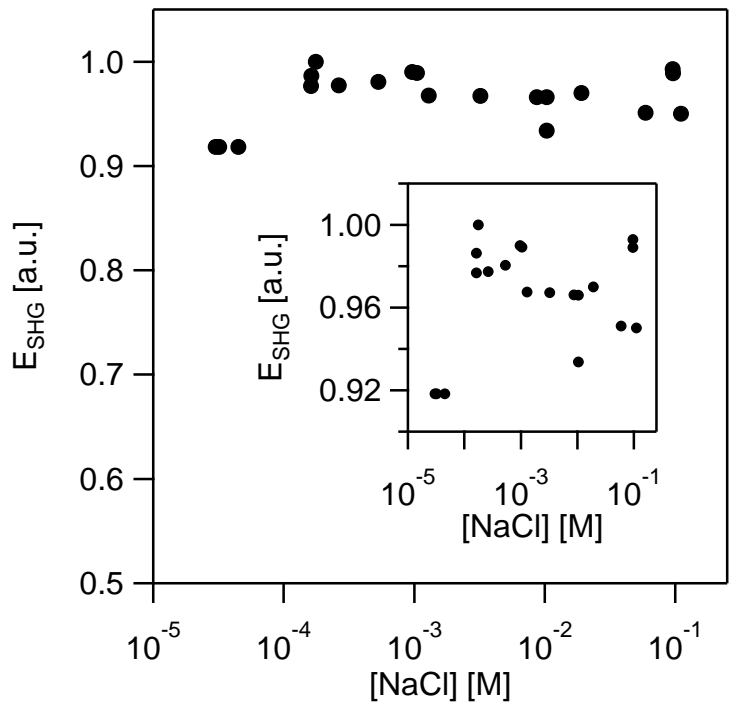

**Figure 2.** SHG E-field as a function of NaCl ionic strength at the $1\bar{1}02$ sapphire/water interface maintained at pH 5.2, the pH for the point of zero charge (PZC). Inset: data replotted with y-axis set to range from 0.90 to 1.02.

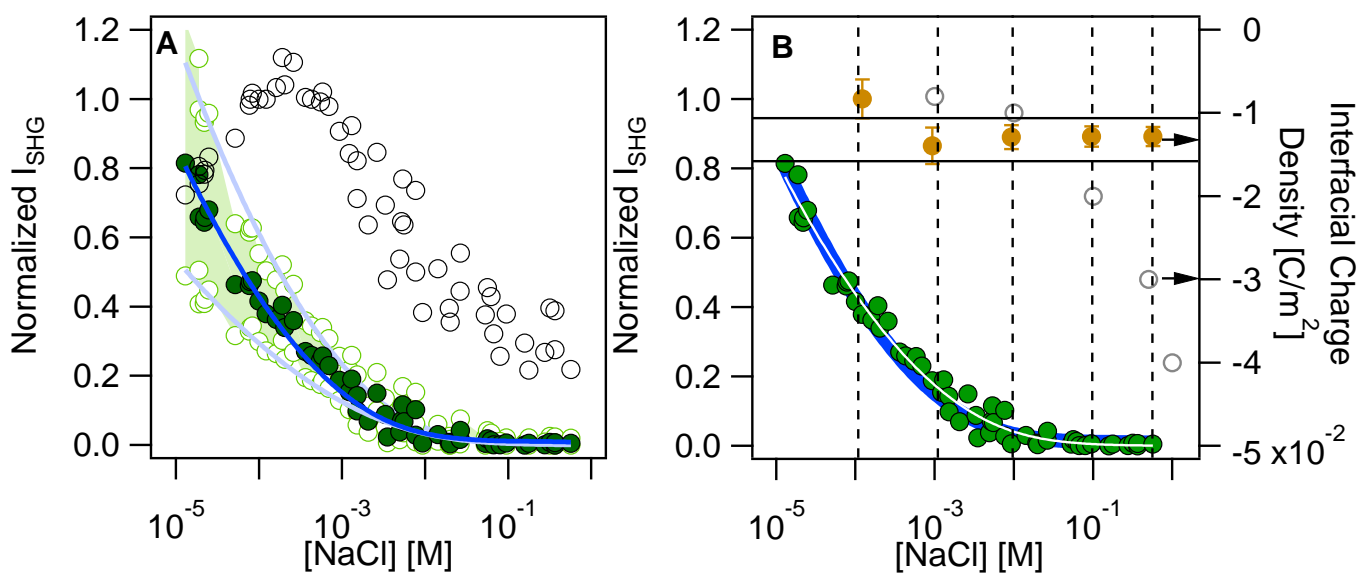

**Figure 3**. (**A**) Observed normalized SHG intensity (empty black circles) and the same corrected via eqn. 2b (filled circles) and upper and low bounds as described in the text (empty green circles) for the fused silica/water interface held at pH 7. Solid lines are fit to eqn. 5. (**B**) Phase matching corrected SHG intensity (left y axis) with Gouy-Chapman model fit (solid white line) results (filled orange circles, right y-axis) and literature data values (empty circles, right y axis) for the surface charge density obtained between the lowest salt concentration and higher concentrations indicated by the dashed vertical lines. Solid blue line indicates model result for salt concentration-dependent surface charge density as described in the main text. Horizontal lines indicate upper and lower bound of surface charge density from literature data in the <10 mM salt regime.

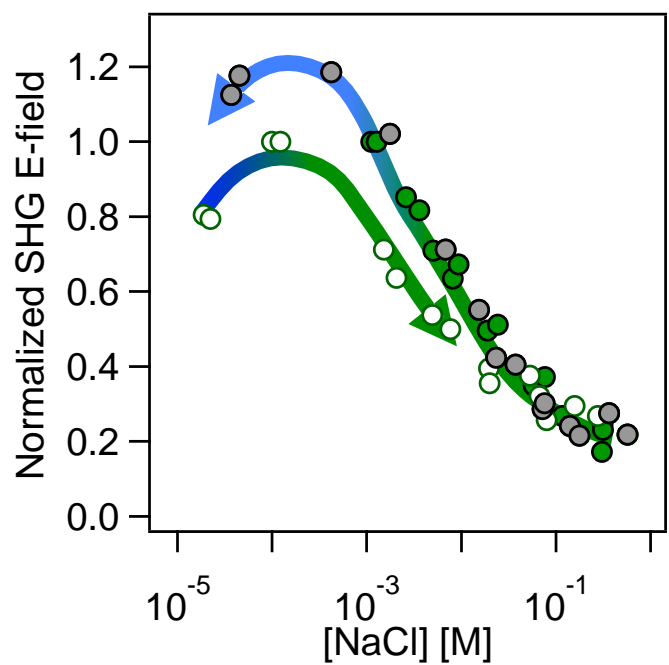

**Figure 4**. Reversibility study screening surface charges at the fused silica/water interface held at pH 7 by first increasing the salt concentration from low to high (empty circles) followed by return to low salt concentration (filled grey circles), as indicated by the arrows. Filled green circles indicate result from a separate charge screening experiment carried out by salt screening from low to high salt concentration following initial exposure of fused silica to 300 mM NaCl from three hours.

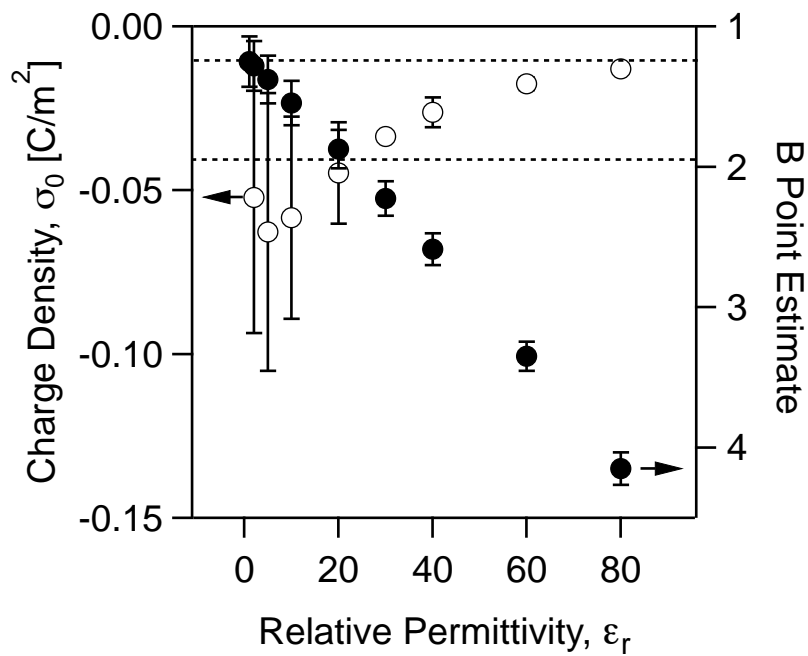

**Figure 5**. Surface charge densities (empty circles) from the Gouy-Chapman model fits to the phase matching-corrected SHG data using relative permittivity values between 1 and 80 along with point estimates for the B parameter (filled circles). Horizontal lines indicate upper and lower bound of surface charge density from literature data in the <1 M salt regime. Please see text for details.